\ifx\mnmacrosloaded\undefined \input mn\fi
%

\newif\ifAMStwofonts
\ifCUPmtplainloaded \else
  \NewTextAlphabet{textbfit} {cmbxti10} {}
  \NewTextAlphabet{textbfss} {cmssbx10} {}
  \NewMathAlphabet{mathbfit} {cmbxti10} {} 
  \NewMathAlphabet{mathbfss} {cmssbx10} {} 
  \ifAMStwofonts
    \NewSymbolFont{upmath} {eurm10}
    \NewSymbolFont{AMSa} {msam10}
    \NewMathSymbol{\upi}     {0}{upmath}{19}
    \NewMathSymbol{\umu}     {0}{upmath}{16}
    \NewMathSymbol{\upartial}{0}{upmath}{40}
    \NewMathSymbol{\leqslant}{3}{AMSa}{36}
    \NewMathSymbol{\geqslant}{3}{AMSa}{3E}

    \let\leq=\leqslant \let\le=\leqslant
     \let\ge=\geqslant
  \else
    \def\umu{\mu}
    \def\upi{\pi}
    \def\upartial{\partial}
  \fi
\fi

\input psfig.tex


\pageoffset{-1.0pc}{0pc}

\loadboldmathnames


\pagerange{000--000}    
\pubyear{1996}
\volume{000}

\def\la{\mathrel{\hbox{\rlap{\hbox{\lower4pt\hbox{$\sim$}}}\hbox{$<$}}}}
\def\ga{\mathrel{\hbox{\rlap{\hbox{\lower4pt\hbox{$\sim$}}}\hbox{$>$}}}}

\def\ax{$\alpha_{\rm x}$}
\def\ea{et al.\ }

\begintopmatter  

\title{Are the X-ray spectra of flat-spectrum radio quasars and BL Lacertae
objects different?} 
\author{Paolo Padovani$^1$, Paolo Giommi$^2$, and Fabrizio Fiore$^{2,3}$}

\affiliation{$^1$ Dipartimento di Fisica, II Universit\`a di Roma ``Tor
Vergata'', Via della Ricerca Scientifica 1, I-00133 Roma, Italy}
\medskip
\affiliation{$^2$ SAX, Science Data Center, ASI, Viale Regina Margherita 202,
I-00198 Roma, Italy} 
\medskip
\affiliation{$^3$ Osservatorio Astronomico di Roma, Via dell'Osservatorio 5,
I-00040 Monteporzio Catone (Roma), Italy}

\shortauthor{P. Padovani, P. Giommi, and F. Fiore}

\shorttitle{X-ray spectra of FSRQ and BL Lacs}


\acceptedline{Accepted . Received ; in original form }

\abstract {We study the X-ray spectra of 114 flat-spectrum radio quasars
(FSRQ) using the hardness ratios as given in the WGA catalogue of {\it ROSAT}
sources. This sample includes all WGA FSRQ with high-quality data and
comprises about 20 per cent of presently known such objects, which makes this 
the largest FSRQ sample ever studied in the X-ray band. We find that FSRQ
have a distribution of energy spectral indices ranging between 0 and 3 with a
mean value \ax~$\sim 1$. This is consistent with that of low-energy cutoff
BL Lacs (LBL; \ax~$\sim 1.1$), generally found in radio surveys, but
significantly different from that of high-energy cutoff BL Lacs (HBL),
normally selected in the X-ray band, which display steeper X-ray spectra
(\ax~$\sim 1.5$). The shape of the optical-to-X-ray continuum is concave (that
is \ax~$< \alpha_{\rm ox}$) for the majority of FSRQ, as found for LBL, 
supporting a dominance of inverse Compton emission in the X-ray band in most 
objects. Our
results are at odds with previous studies of the X-ray spectra of FSRQ, which
were however plagued by low spectral resolution and/or small number statistics
and selection effects, and have important implications for the proposed
connections between FSRQ and BL Lacs.} 


\maketitle  


\section{Introduction} 

BL Lacertae objects constitute an extreme and relatively rare type of active
galactic nuclei (AGN), characterized by high luminosity, rapid variability,
high ($> 3$ per cent) optical polarization, radio core-dominance, superluminal
velocities, and almost complete lack of emission lines (e.g. Kollgaard 1994;
Urry \& Padovani 1995). Furthermore, the broad-band emission in these objects,
which extends from the radio all the way up to the gamma-ray band, is most
probably dominated by non-thermal processes, undiluted by the thermal emission
present in most other AGN (e.g. Bregman 1990). 

Similar properties (apart from the presence of a UV ``bump'' in some cases),
are also displayed by a subclass of radio-loud quasars, whose spectra
therefore show, by definition, strong, broad emission lines. These 
have typical equivalent widths much larger than 5 \AA, the value below
which objects are generally classified as BL Lacs (e.g. Stickel \ea 1991; 
Stocke \ea 1991). These quasars are defined variously as Optically
Violently Variable (OVV) quasars, Highly Polarized Quasars (HPQ),
Core-Dominated Quasars (CDQ), or Flat-Spectrum Radio Quasars (FSRQ), the 
latter indicating a radio spectral index $\alpha_{\rm r} < 0.5$ ($f_{\nu}
\propto \nu^{-\alpha}$) at a few GHz (e.g. Urry \& Padovani 1995). Although
these names reflect different empirical definitions, growing evidence suggests 
that these various classes coincide, that is FSRQ tend to show rapid 
variability, high polarization, and radio structures dominated by compact 
radio cores, and vice versa (e.g. Fugmann 1988; Impey \& Tapia 1990). 
BL Lacs and FSRQ are often collectively called blazars. 

The relation between the two classes, if one exists, is not clear. Padovani
(1992), based mostly on their different isotropic properties, has argued that
BL Lacs and FSRQ represent examples of similar relativistic phenomena hosted
by radio galaxies of different power, BL Lacs being associated with
Fanaroff-Riley type I (i.e. low luminosity: Fanaroff \& Riley 1974) radio
galaxies, and FSRQ being associated with FR II (i.e. high luminosity) radio
galaxies (see also Urry \& Padovani 1995 and references therein). A tighter
connection has also been proposed, be it through evolution (e.g. Vagnetti,
Giallongo \& Cavaliere 1991) or gravitational lensing (e.g. Ostriker \&
Vietri 1985). 

An important part in the study of the relation between the two classes is
played by the shape of the X-ray spectra. For example, if the majority of BL
Lacs were actually gravitationally microlensed radio quasars (Ostriker and
Vietri 1985), the X-ray spectra of BL Lacs and FSRQ should probably be 
indistinguishable (see also Worrall \& Wilkes 1990). Similarly, if FSRQ evolve
into BL Lacs through an increase with cosmic time of the Lorentz factor
(Vagnetti \ea 1991), this should affect their X-ray spectra. 

A detailed comparison of the X-ray spectra of BL Lacs and FSRQ is then clearly
important in this respect, as it can further constrain the proposed connection
(or lack of) between them and, more generally, shed light on the emission
mechanisms responsible for the X-ray emission in both classes. Although
previous results have suggested FSRQ to have harder X-ray spectra than BL
Lacs, these were actually strongly affected by low spectral resolution and/or
small statistics and selection biases. Worrall \& Wilkes (1990) analyzed the
{\it Einstein} IPC spectra of 31 FSRQ (including 12 HPQ) and 23 radio-selected
BL Lacs (RBL). It was found that, letting $N_{\rm H}$ free to vary in the
single fits, FSRQ had $\alpha_{\rm x} \sim 0.5$ while RBL 
had $\alpha_{\rm x} \sim 1$ (where \ax~denotes the energy index), which was
interpreted in terms of a different mixing of beamed and unbeamed X-ray 
emission between the two classes. The derived spectral indices, however, had
very large uncertainties due to the poor spectral resolution of the IPC
experiment. More recently, Brunner \ea (1994) have studied the {\it ROSAT}
X-ray spectra of a complete (flux limited) sample of 13 flat-spectrum sources
extracted from 
the S5 radio catalogue, 8 FSRQ and 5 BL Lacs. They found $\alpha_{\rm x} =
0.59\pm 0.19$ for the quasars and $\alpha_{\rm x} = 1.36\pm 0.27$ for the BL
Lacs. The two classes, however, had largely different redshift distributions,
FSRQ being at higher redshift, with half the FSRQ at $z > 1.5$ and only two 
FSRQ in the BL Lac redshift range. The X-ray spectrum of radio-loud quasars is
likely to flatten with redshift (Schartel \ea 1996; Fiore \& Elvis 1995; Elvis
\ea 1994; Bechtold \ea 1994), so this redshift difference is going to bias the
comparison between 
the two classes. Finally, Urry \ea (1996) have compared the X-ray spectral
index distribution of the Brunner \ea (1994) FSRQ with that of the 1-Jy sample
of BL Lacs. Again, the \ax~distributions of the two classes are significantly
different, but so are the redshift distributions, with still only two FSRQ in
the BL Lac redshift range. 

The purpose of this paper is to analyze the X-ray spectra of all FSRQ observed
(as pointed or serendipitous sources) by {\it ROSAT} and compare them to those
of 85 BL Lacs, studied in a previous paper (Padovani \& Giommi 1996, hereafter
Paper I). Our data have been taken from the WGA catalogue (White, Giommi \&
Angelini 1994), a large list of X-ray sources generated from all the {\it
ROSAT} PSPC pointed observations, with which it is now possible to study the
X-ray properties of large numbers of objects in an homogeneous and relatively
simple way. The selection of the objects was done by cross-correlating the
WGA catalogue with various optical and radio catalogues. This resulted in 225
observations of 114 FSRQ, which make up about 18 per cent of known FSRQ (see
Section 2.2). 

The comparison between FSRQ and BL Lac X-ray spectra can therefore be
performed for the first time in a way which satisfies all of the following
important requirements: large number statistics, reasonably good X-ray spectral 
resolution, large redshift range and in particular good overlap between the
redshift distributions of the two classes (see Section 2.2). The structure of
the paper is as follows: Section 2 describes the BL Lac and FSRQ samples used
in this work, Section 3 deals with the observational data and their analysis,
Section 4 studies the X-ray spectral properties of FSRQ while Section 5
discusses our results and presents our conclusions. Throughout this paper
spectral indices are written $f_{\nu} \propto \nu^{-\alpha}$. 

\section{The Samples}

\subsection{BL Lacs}

The BL Lac sample was put together in Paper I by cross-correlating the first
revision of the {\it ROSAT} WGA catalogue with a recent BL Lac catalogue
(Padovani \& Giommi 1995b). It includes 85 sources, which correspond to $\sim
50$ per cent of confirmed BL Lacs presently known. BL Lacs appear to come in
two types (e.g. Stocke \ea 1985; Ledden \& O'Dell 1985; Giommi, Ansari \&
Micol 1995; Padovani \& Giommi 1995a): those with a peak in their broad-band
spectrum at relatively low (infrared/optical) energies (LBL), mostly found in
radio surveys, and those with a peak at relatively high (ultraviolet/X-ray)
frequencies (HBL), which are typically selected in the X-ray band. The 
dividing line between the two classes is at a ratio $f_{\rm x}/f_{\rm r} \sim
10^{-11.5}$ (with X-ray fluxes in the 0.3 -- 3.5 keV range in units of erg 
cm$^{-2}$ s$^{-1}$ and radio fluxes at 5 GHz in janskys), HBL having an
X-ray-to-radio flux ratio above this value (Paper I). The properties of the
two subclasses are somewhat different, those of LBL being more extreme
(Kollgaard 1994; Urry \& Padovani 1995). Giommi \ea (1995) and Padovani \&
Giommi (1995a) have argued that there may be only one population of objects,
characterized by a wide range of peak energies, and that the existence of two 
classes simply reflects the different selection criteria of radio and 
X-ray surveys. Our BL Lac sample includes 58 HBL and 27 LBL (see Paper I). 

\subsection{Flat-spectrum radio quasars}

The FSRQ were selected by cross-correlating the first revision of the WGA
catalogue (restricting ourselves to sources with quality flag $\ge 5$, which
excludes problematic detections) with a variety of optical and radio
catalogues, including the V\'eron-Cetty \& V\'eron (1993) and Hewitt \&
Burbidge (1993) quasar catalogues, and the 1-Jy (Stickel, Meisenheimer \&
K\"uhr 1994) and S4 (Stickel \& K\"uhr 1994) radio catalogues. All objects in
this study are ``bona fide'' quasars, that is objects whose optical
spectrum has a strong nonstellar component and shows broad lines. Broad-line
radio galaxies (with flat radio spectra) are then also included (see e.g. Urry
\& Padovani 1995). Two sources classified as quasars in the Parkes catalogue
on the basis of their optical ``stellar'' appearance and for which no redshift
was available were excluded. 3C 111, known to be strongly absorbed (Nandra \&
Pounds 1994), and NRAO 140, for which there is evidence of variable absorption
(Turner \ea 1995), were 
also excluded. Excluding from our analysis observations with values of
hardness ratios relative to their uncertainties $< 3.5$ (see Section 3), 
we are then left with 225 observations of 114 FSRQ, which make 
up about 18 per cent of the FSRQ listed in the V\'eron-Cetty \& V\'eron (1993)
catalogue. This represents the largest number of FSRQ for which homogeneous
X-ray spectral information is available and the largest FSRQ sample ever
studied at X-ray frequencies. The flat radio spectrum classification
($\alpha_{\rm r} < 0.5$) refers to the spectrum at a few GHz and has been
made on the basis of radio data provided by a variety of radio catalogues. 
Fifty-eight objects, that is about 50 per cent of the sample, belong to the 1-Jy
radio catalogue (Stickel \ea 1994), an all-sky flux-limited sample which
includes 527 sources with 5 GHz fluxes $\ge 1$ Jy, 222 of which are FSRQ.
Twenty-four objects, that is about 20 per cent of the sample, belong to the 
2-Jy radio catalogue (Wall \& Peacock 1985), an all-sky flux-limited sample
which includes 233 sources with 2.7 GHz fluxes $\ge 2$ Jy, 51 of which are
FSRQ. 

\beginfigure{1}
\psfig{file=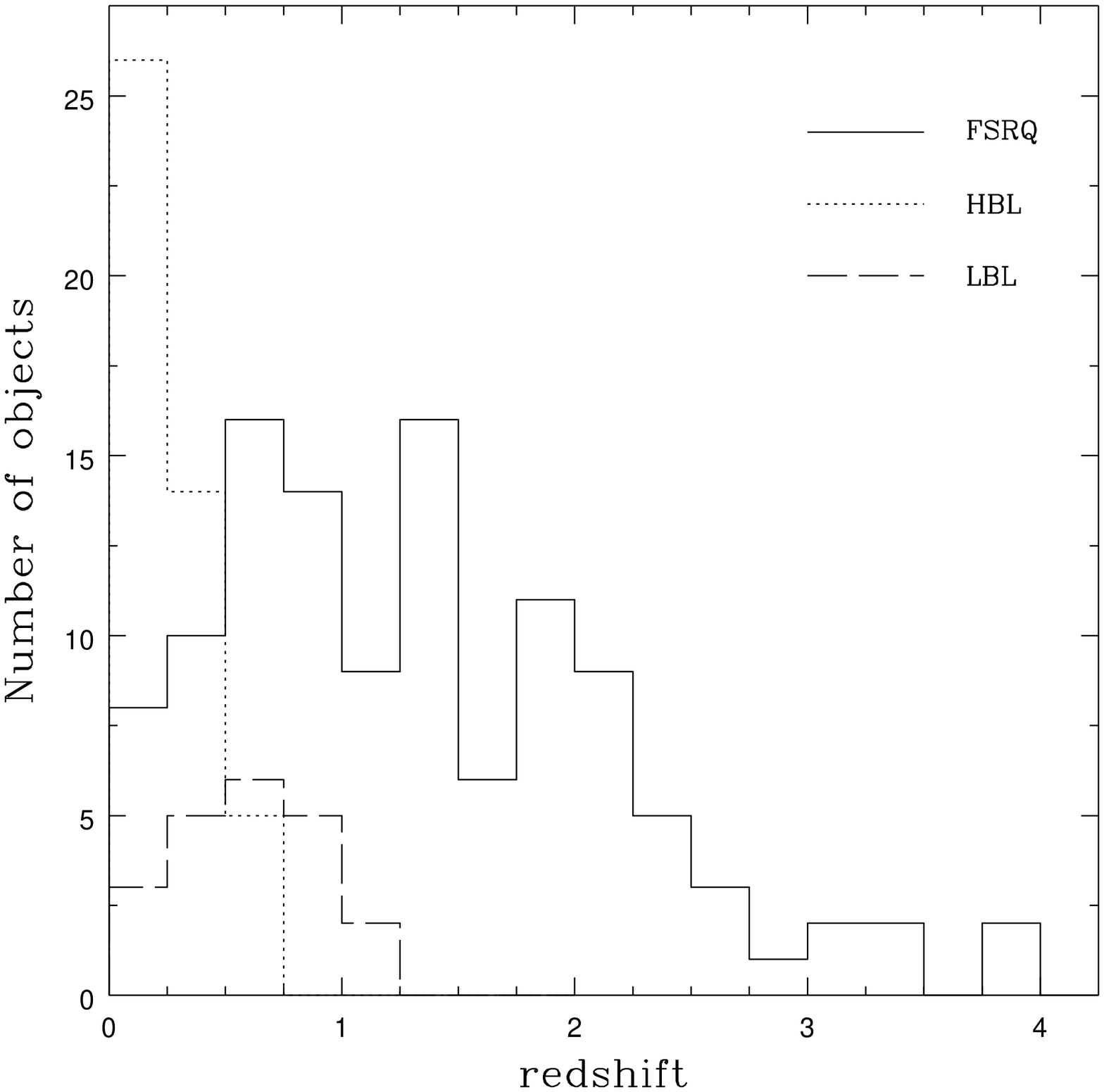,height=8.5truecm,width=8.5truecm}
\caption{{\bf Figure 1.} The redshift distribution of the FSRQ sample (solid
line), compared to that of the HBL (BL Lacs with $f_{\rm x}/f_{\rm r} \ge
10^{-11.5}$: dotted line) and the LBL (BL Lacs with $f_{\rm x}/f_{\rm r} <
10^{-11.5}$: dashed line) studied in Paper I. BL Lacs without redshift 
information are not included.} 
\endfigure

Fig. 1 shows the redshift distribution of the FSRQ sample, which extends up to
$z \sim 4$ with $\langle z \rangle = 1.33\pm0.08$ (here and in the following
we give the standard deviation of the mean). The figure shows also the
redshift distributions of the LBL and HBL objects with redshift information
($\sim 80$ per cent of the BL Lac sample): the overlap between FSRQ and LBL is
substantial, with 50 FSRQ (44 per cent of the sample) having $z \le 1.048$,
the maximum BL Lac redshift. The detailed redshift dependence of the X-ray
spectral indices of FSRQ will be studied elsewhere (Fiore et al., in
preparation) but we can anticipate that a correlation between \ax~and redshift
seems to be present only when objects with $z \ga 2$ are included. In the
following, then, we will compare the X-ray properties of BL Lacs to those
of the whole FSRQ sample and, to exclude the likely redshift dependence, to
those of the $z \le 2$ subsample (90 sources, 79 per cent of the sample) and
the $z \le 1.048$ subsample (50 sources, 44 per cent of the sample). While the
former subsample represents a compromise between good number statistics and a
relatively low mean redshift, the latter has a redshift distribution
indistinguishable from that of the LBL sample according to a
Kolmogorov-Smirnov (KS) test. 

\section{Data Analysis}

Spectral indices for the WGA FSRQ were obtained as described in Paper I for
the BL Lacs. Namely, we derived hardness ($HR$) and softness ($SR$) ratios
from the count rates given in the catalogue in the $0.1-0.4$ keV range (soft
band: $S$), $0.4-0.86$ keV range (mid band: $M$) and $0.87-2.0$ keV range
(hard band: $H$). The count rates were then combined to construct one $SR = S/M$
and two HRs, $HR_1 = H/M$ and $HR_2 = H/(M+S)$. Initially we converted the
hardness ratios into energy spectral indices both assuming Galactic $N_{\rm
H}$ derived from 21 cm measurements (Stark et al. 1992; Shafer et al., private
communication) and with $N_{\rm H}$ derived from the softness ratio through an
iterative procedure. The spectral indices derived from the two approaches were
found to be similar and well correlated in the $0.4-2.0$ keV range, while
considerable scatter was present when the total {\it ROSAT} band was
considered, with a few objects having $\alpha_{\rm x,NH}$ significantly
different from $\alpha_{\rm x,Galactic~NH}$, suggestive of a change of the
spectrum at lower energies. In this paper, as in Paper I, we will adopt as
X-ray spectral index the value obtained with $N_{\rm H}$ fixed to the Galactic
value, which is better determined than the $N_{\rm H}$ derived from the
softness ratios, and in the $0.4-2.0$ keV range, i.e. the mid to hard band. 

The choice of this energy range stems from two facts: first, our method {\it
assumes} a single power-law; some objects, however, show evidence of low-energy
absorption (Fiore et al., in preparation), so this assumption might be
incorrect in the whole {\it ROSAT} range while it is probably more acceptable
in the narrower $0.4-2.0$ keV range (as shown by the similarity of the
spectral indices obtained with Galactic $N_{\rm H}$ and $N_{\rm H}$ derived
from the softness ratio); second, the method used in the WGA catalogue to
estimate the source intensity uses the counts detected in a box whose size
optimises the signal to noise (S/N) ratio. This size is calculated assuming an
average point spread function (PSF) that is too sharp for very soft photons.
For weak sources near the field center the adopted box size is small and
includes an energy-dependent fraction of the source
photons causing an underestimation of the counts in the soft band. 
The effects of the dependence of energy response with off-axis angle have been
taken into account using different conversion matrices in five different
off-axis ranges: $0-20$, $20-30$, $30-40$, $40-50$, and $50-60$ arcminutes.
Note that this method of estimating spectral indices is quite robust, as shown
in Paper I, and particularly suitable for the determination of the X-ray
spectral slope distribution for large samples of objects (Giommi et al., in
preparation). 

Errors on the spectral indices ($1 \sigma$) were derived from the
uncertainties on the hardness ratios. We included in our analysis only
observations with values of hardness ratios relative to their uncertainties $>
3.5$, which corresponds roughly to $1\sigma$ errors on $\alpha_{\rm x} \la
0.5$ (Giommi et al., in preparation). 

A few FSRQ have been repeatedly observed, typically in the course of
multi-frequency campaigns (e.g. 3C 273, 3C 279) and in general many sources
have more than one observation. Inclusion of all these data in statistical
studies (means, correlations, KS tests etc.) would
clearly bias them, as the results would be weighed towards the sources with
the largest number of observations. Therefore, although we plot in some figures
the data referring to all 225 observations, in the statistical tests we have
kept only one observation (and therefore one spectral index) per object,
selected on a one-to-one basis taking the best combination of offset from the
field center and S/N ratio. When more than two observations with comparable
offset and S/N ratios were available, we selected as most representative the
one with \ax~closest to the mean value for that source. 

\section{The X-ray spectra}

The X-ray spectral index distribution of the FSRQ in our sample is shown in
Fig. 2. The mean value is $\langle \alpha_{\rm x,FSRQ}
\rangle = 1.04\pm0.05$ (see also Table 1). A KS test shows that this
distribution and that of the 85 BL 
Lacs studied in Paper I ($\langle \alpha_{\rm x} \rangle = 1.37\pm0.05$) are
different at the 99.99 per cent level. Even if one compares the BL Lacs to the
$z \le 2$ and $z \le 1.048$ FSRQ subsamples (which have slightly steeper
averages $\langle \alpha_{\rm x} \rangle = 1.09\pm0.05$ and $\langle
\alpha_{\rm x} \rangle = 1.16\pm0.06$: see Table 1), the BL Lac and FSRQ
\ax~distributions are still different at the 99.8 and 97.1 per cent level
respectively. However, this is clearly due to the 
relative steepness of the X-ray spectra of HBL ($\langle \alpha_{\rm x,HBL}
\rangle = 1.52\pm 0.06$). In fact, the \ax~distribution of the 27 LBL, 
for which $\langle \alpha_{\rm x,LBL} \rangle = 1.06\pm 0.09$, is {\it fully
consistent} with that of the whole FSRQ sample (see Fig. 2) and the two lower
redshift subsamples, according to a KS test. 

\begintable{1}
  \caption{{\bf Table 1.} Mean values.} 
  \halign{#\hfil&\quad \hfil#\quad& \hfil#\hfil\quad&
          \hfil#\quad& #\hfil\cr
  Sample  & N & $\langle \alpha_{\rm x} \rangle$ & $\langle \alpha_{\rm x} -
  \alpha_{\rm ox} \rangle$&   Notes \cr 
  FSRQ                & 114 & $1.04\pm0.05$ & $-0.28\pm0.05$ & This work\cr
  FSRQ ($z \le 2$)    &  90 & $1.09\pm0.05$ & $-0.22\pm0.05$ & This work\cr
  FSRQ ($z \le 1.048$)&  50 & $1.16\pm0.06$ & $-0.12\pm0.06$ & This work\cr
  All BL Lacs         &  85 & $1.37\pm0.05$ &  $0.17\pm0.06$ & Paper I\cr
  LBL                 &  27 & $1.06\pm0.09$ & $-0.33\pm0.09$ & Paper I\cr
  HBL                 &  58 & $1.52\pm0.06$ &  $0.40\pm0.04$ & Paper I\cr
  }
  \tabletext{Mean \ax~and $\alpha_{\rm x} - \alpha_{\rm ox}$ values for 
    the samples discussed in this paper. N is the number of objects in the 
    various samples.}
\endtable

\beginfigure{2}
\psfig{file=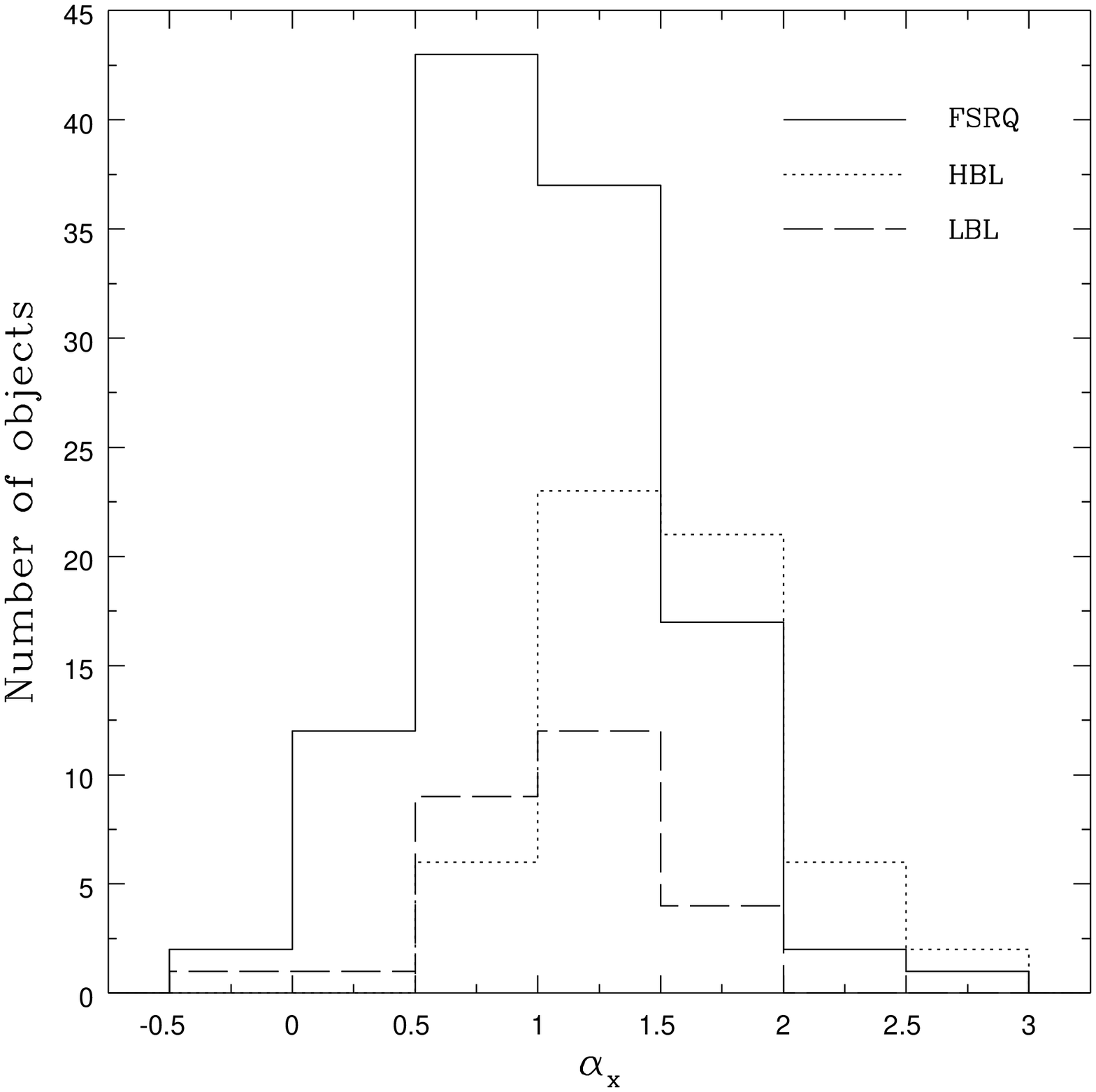,height=8.5truecm,width=8.5truecm}
\caption{{\bf Figure 2.} The X-ray spectral index distribution of the FSRQ in
our sample (solid line), compared to that of the HBL (BL Lacs with $f_{\rm
x}/f_{\rm r} \ge 10^{-11.5}$: dotted line) and the LBL (BL Lacs with $f_{\rm
x}/f_{\rm r} < 10^{-11.5}$: dashed line) studied in Paper I. Only one X-ray
spectrum per object is used, as described in Section 2. The distributions of
FSRQ and LBL are indistinguishable according to a KS test.} 
\endfigure

Paper I showed that the X-ray spectral slope of HBLs was strongly
correlated with the effective optical-X-ray spectral index $\alpha_{\rm ox}$
(evaluated between the rest-frame frequencies of 5000 \AA~and 1 keV),
indicative of a common origin for the optical/X-ray emission in those objects.
Moreover, all but one HBL had an overall convex spectrum, that is $\alpha_{\rm
x} \ge \alpha_{\rm ox}$ (within the errors). On the other hand, no such
correlation was present for LBL, the majority of which had $\alpha_{\rm x} <
\alpha_{\rm ox}$, that is a concave optical-X-ray spectrum. It is therefore
interesting to see how FSRQ behave in this respect. Fig. 3 plots \ax~versus
$\alpha_{\rm ox}$ for FSRQ and LBL (the latter data from Paper I; HBL are not 
included to improve legibility). Optical
fluxes for FSRQ were derived from the V magnitudes included in the catalogues
listed in Section 2.2, correcting for Galactic absorption following Giommi \ea
(1995), while 1 keV fluxes have been obtained from the {\it ROSAT} counts and
the derived spectral indices and have also been corrected for Galactic
absorption. The $k$-correction has been derived assuming an optical
index $\alpha_{\rm o} = 1$, a value which is intermediate between $\alpha_{\rm
o} = 1.4$, reported by Ghisellini \ea (1986) for 6 blazars with strong
emission lines, and $\alpha_{\rm o} = 0.5$, obtained by Baker \& Hunstead
(1995) for 13 core-dominated quasars. Note that optical fluxes are not
simultaneous with X-ray data which, given the strong optical/X-ray variability
of FSRQ, will certainly introduce a scatter: a variation $\Delta V$ in
magnitude, for example, translates into a change $\Delta \alpha_{\rm ox} =
0.15 \Delta V$. 

\beginfigure{3}
\psfig{file=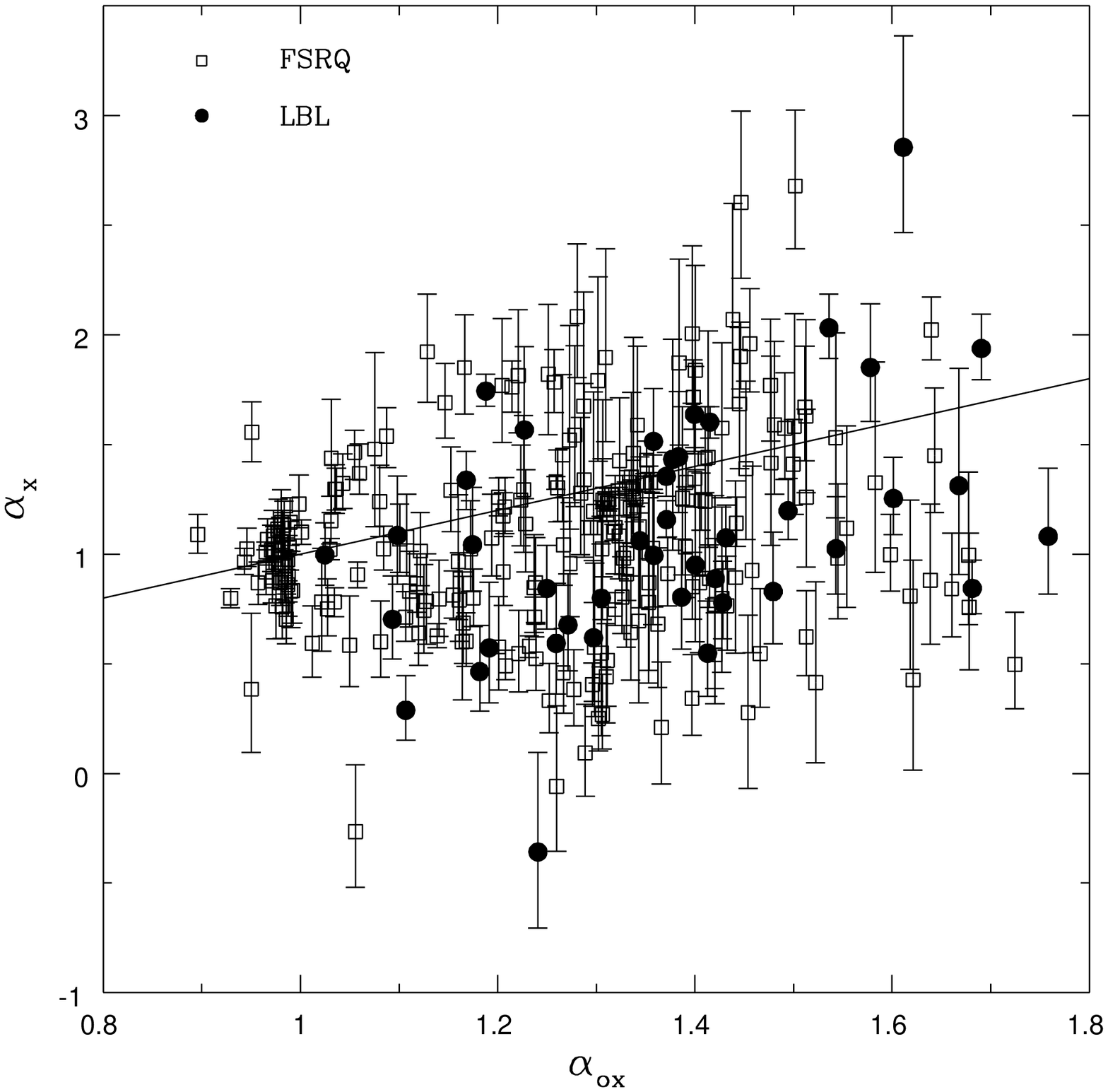,height=8.5truecm,width=8.5truecm}
\caption{{\bf Figure 3.} The X-ray spectral index versus $\alpha_{\rm ox}$ for
the FSRQ in our sample and the LBL studied in Paper I. Filled points indicate
LBL (BL Lacs with $f_{\rm x}/f_{\rm r} < 10^{-11.5}$), while open squares
indicate FSRQ. The solid line represents the locus of points having
\ax~$=\alpha_{\rm ox}$. Error bars represent $1\sigma$ errors.} 
\endfigure

The data presented in Fig. 3 show that, as is the case for LBL, there is no
correlation between \ax~and $\alpha_{\rm ox}$ for FSRQ. Moreover, the
distribution of the two classes on the \ax~- $\alpha_{\rm ox}$ plane is
indistinguishable according to a two-dimensional KS test (Fasano \&
Franceschini 1987). It then follows that the $\alpha_{\rm x} - \alpha_{\rm ox}$
distributions of the two classes cannot be that different, as illustrated by
Fig. 4. In fact, $\langle \alpha_{\rm x} - \alpha_{\rm ox} \rangle =
-0.33\pm0.09$ for LBL (Paper I) and $-0.28\pm0.05$ for FSRQ, with the latter
value changing to $-0.22\pm0.05$ for $z \leq 2$ and to $-0.12\pm0.06$ for $z
\le 1.048$ (Table 1). According to a Student's t-test, the mean values for BL
Lacs and FSRQ are not significantly different in all cases, although barely so
for the $z \le 1.048$ FSRQ subsample (which has a value of $\langle
\alpha_{\rm x} - \alpha_{\rm ox} \rangle$ different from that of LBL at the 94
per cent level). For comparison,
Fig. 4 shows also the $\alpha_{\rm x} - \alpha_{\rm ox}$ distribution for HBL,
from Paper I, clearly different from that of both FSRQ and LBL, with $\langle
\alpha_{\rm x} - \alpha_{\rm ox} \rangle = 0.40\pm0.04$. This reflects the
different distributions of HBL on the \ax~- $\alpha_{\rm ox}$ plane (see Paper
I). Note also that only 30 per cent of FSRQ have convex optical-to-X-ray
continua (i.e. $\alpha_{\rm x} > \alpha_{\rm ox}$), to be compared with 25 per
cent of LBL (and 98 per cent of HBL; see Paper I). Assuming that the errors
associated with $\alpha_{\rm x} - \alpha_{\rm ox}$ equal those on \ax, a
clearly conservative assumption, it turns out that only 13 FSRQ (or 11 per
cent of the sample) have $\alpha_{\rm x} > \alpha_{\rm ox}$ at the $\ga
2\sigma$ level. This is similar to the fraction of LBL (15 per cent; Paper I)
having convex spectra at the same significance level. 

\beginfigure{4}
\psfig{file=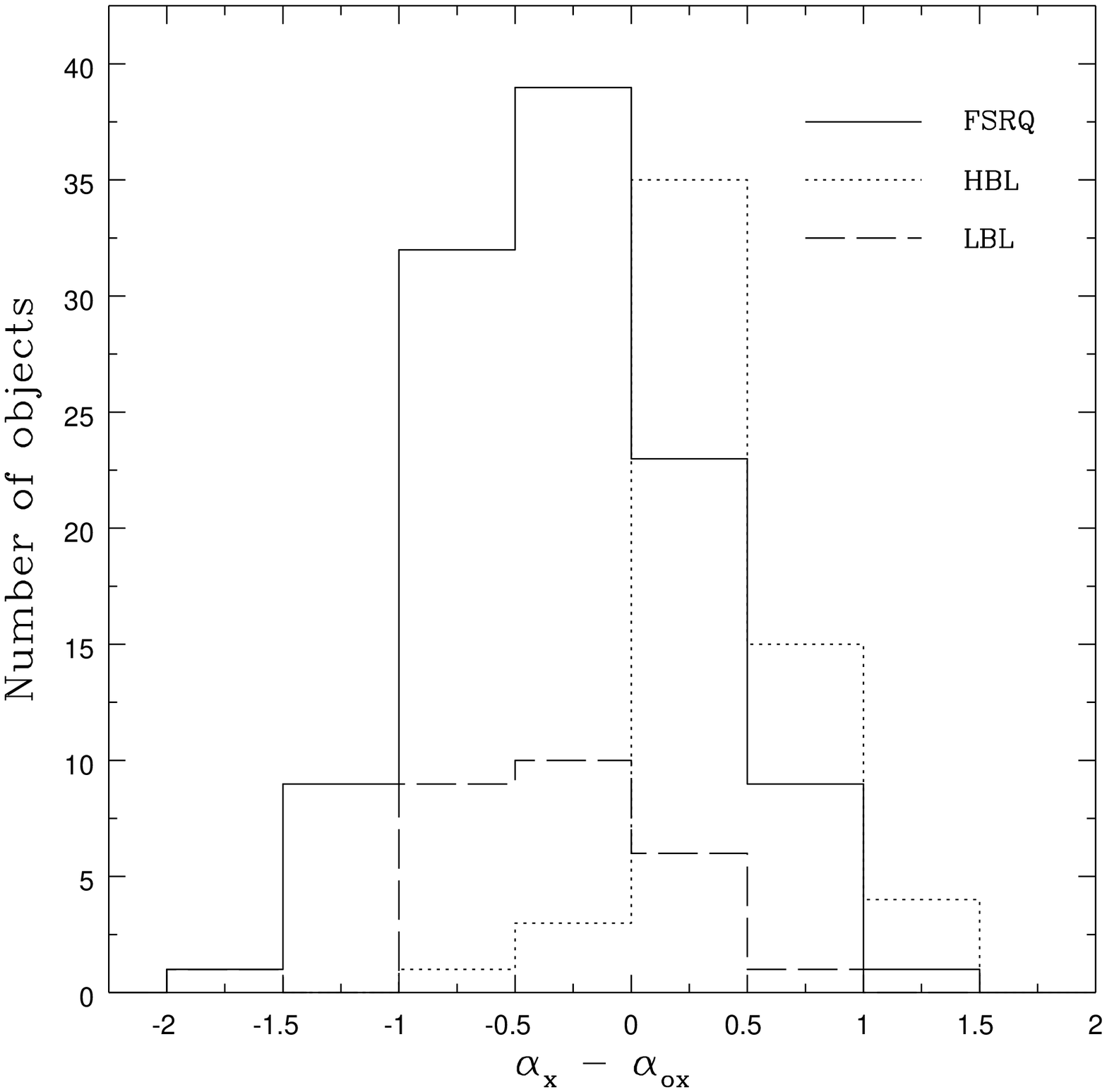,height=8.5truecm,width=8.5truecm}
\caption{{\bf Figure 4.} The distribution of \ax~minus $\alpha_{\rm ox}$ 
for the FSRQ in our sample (solid line), compared to that of the HBL (BL Lacs
with $f_{\rm x}/f_{\rm r} \ge 10^{-11.5}$: dotted line) and the LBL (BL Lacs
with $f_{\rm x}/f_{\rm r} < 10^{-11.5}$: dashed line) studied in Paper I. Only
one X-ray spectrum per object is used, as described in Section 2. The
distributions of FSRQ and LBL are indistinguishable according to a KS test.} 
\endfigure

\section{Discussion and conclusions}

The data presented in this paper indicate that the
X-ray spectra of FSRQ are similar to those of LBL. Moreover, the two classes
have also a similar distribution on the \ax~- $\alpha_{\rm ox}$ plane and
similar mean values of $\alpha_{\rm x} - \alpha_{\rm ox}$, with most sources
exhibiting a concave optical/X-ray continuum. On the other hand, the X-ray
spectra of FSRQ are, on average, flatter than those of HBL, which moreover
display an overall convex spectrum.

Paper I showed that the X-ray spectra of the two BL Lac classes were different,
HBL having steeper spectra. This and other results tied in with the different
multifrequency spectra and suggested that while the flatter X-ray emission of
LBL was dominated by inverse Compton emission, as the synchrotron break in
these objects is in the infrared/optical band, that of HBL was an extension of
the synchrotron emission also responsible for the lower energy continuum.

The results of the present paper, therefore, favour a dominance of inverse
Compton emission in FSRQ as well. (This does not necessarily mean that the
{\it ROSAT} band is dominated by pure Compton emission in {\it all} FSRQ, as
some objects do display a steep spectrum, most probably due to some other soft
component. One would expect, however, that FSRQ as a class had a flatter
spectral slope at harder X-ray energies such as those accessible to {\it SAX}  
and {\it ASCA}.) As a corollary, then, one would infer that
most FSRQ should have the peak of their emission at infrared/optical
frequencies and therefore their broad-band spectra should be quite similar to
those of LBL. (A complication here is represented by the possible presence of a
UV ``bump'' in FSRQ, i.e. of an optical thermal component independent of the
non-thermal processes present in BL Lacs, which would also increase $\alpha_{\rm
ox}$ for FSRQ. Given the relatively steep UV spectral indices of CDQ [i.e.
FSRQ; Wills \ea 1995], however, its contribution is probably minor, although
possibly not negligible, in most objects. Notable exceptions, however, exist
and include, for example, 3C 273). 

It has been known for some time that radio-loud quasars and radio-selected BL
Lacs occupy similar regions of the $\alpha_{\rm ro}$ - $\alpha_{\rm ox}$ plane,
where the effective spectral indices are defined in the usual way and
calculated here between the rest-frame frequencies of 5 GHz, 5000 \AA~and 1 keV
(e.g. Stocke \ea 1985; Stocke \ea 1991; Padovani 1992). Fig. 5 shows that to
be the case also when only FSRQ are considered: the distribution on the
$\alpha_{\rm ro}$ - $\alpha_{\rm ox}$ plane for the objects studied in this
paper is largely overlapping with that of LBL and clearly distinct from that of
HBL. Although some FSRQ ``invade'' the HBL region they do so only
marginally. The question of why we do not see a population of FSRQ with
broad-band continua peaked in the X-ray band and therefore similar to those of
HBL is an intriguing open problem which we hope to address in a future paper.

\beginfigure{5}
\psfig{file=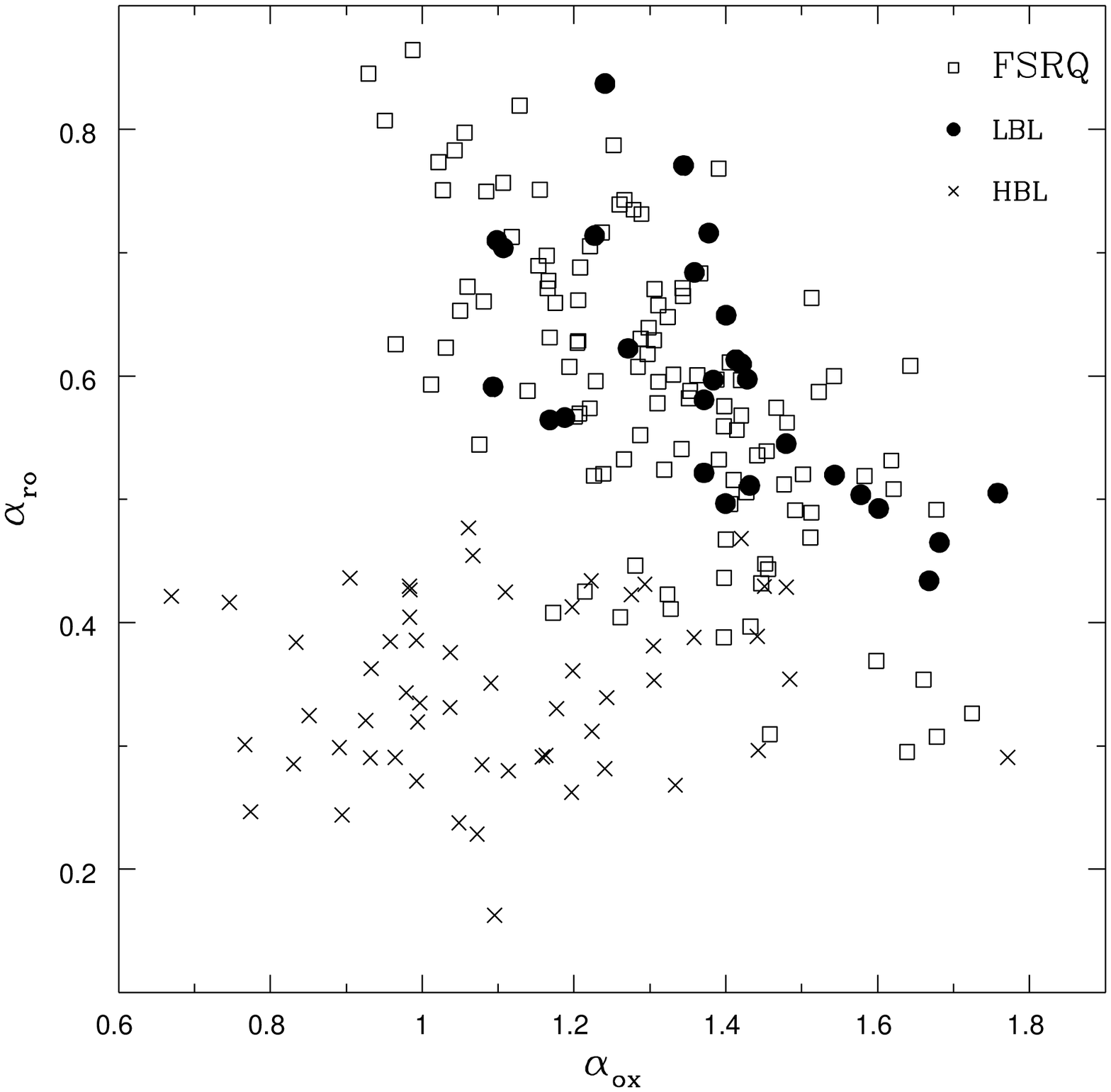,height=8.5truecm,width=8.5truecm}
\caption{{\bf Figure 5.} $\alpha_{\rm ro}$ - $\alpha_{\rm ox}$ plane for the
FSRQ and BL Lacs studied in this paper and in Paper I. The effective spectral
indices are defined in the usual way and calculated between the rest-frame
frequencies of 5 GHz, 5000 \AA~and 1 keV. Crosses indicate HBL (BL Lacs with
$f_{\rm x}/f_{\rm r} \ge 10^{-11.5}$), filled circles indicate LBL (BL Lacs
with $f_{\rm x}/f_{\rm r} < 10^{-11.5}$), while open squares represent FSRQ.
To improve legibility, only one point per object, selected as described in 
Section 2, is plotted. Note that FSRQ and LBL occupy roughly the same region,
distinct from that typical of HBL.} 
\endfigure

The results of our study disagree with those obtained in previous papers,
which found a difference between the X-ray spectra of FSRQ and BL Lacs.
Worrall \& Wilkes (1990), using the {\it Einstein} IPC, derived $\alpha_{\rm
x} \sim 0.5$ for 31 FSRQ and $\alpha_{\rm x} \sim 1$ for 23 radio-selected BL
Lacs. Given that only six of
their FSRQ have $z > 2$, it is unlikely that this 
difference is due to the flattening of \ax~with redshift, previously
discussed. Also, their spectral slopes were affected by very large errors due to
the poor spectral resolution of the IPC experiment (the majority of the 
sources had 90 per cent upper error bars larger than 1, with many objects 
having 90 per cent upper error bars in excess of 2) but it is hard to see how 
this could affect differently the derived spectral slopes for the two classes. 
A possible explanation for this difference could be the following: 
1. as described in Section 3, our spectral indices have been derived assuming 
Galactic $N_{\rm H}$, based on the similarity between spectral slopes derived
under this assumption and those obtained with $N_{\rm H}$ being a free
parameter. In the case of Galactic absorption, Worrall \& Wilkes (1990)
obtained $\alpha_{\rm x} \approx 0.5 - 0.6$ {\it both} for FSRQ and RBL (see
their Fig. 5). Their Table 4 shows that the assumption $N_{\rm H} = N_{\rm H,
gal}$, once the few objects for which this gives poor fits (4 RBL and 4 FSRQ)
are excluded, is acceptable statistically for all classes. Note also that Urry
\ea (1996), in their detailed spectral analysis of {\it ROSAT} data for 28
1-Jy RBL, have found that most objects are well fitted by a single power-law
model with Galactic absorption; 2. it is well known that there is a systematic
difference 
between the IPC and PSPC spectral indices of about $0.4 - 0.5$, the former
being harder (Fiore \ea 1994; Urry \ea 1996; Ciliegi \& Maccacaro 1996). This
has been interpreted in terms of two possible effects: 1. systematic errors in
the calibration of both instruments (Fiore \ea 1994); 2. a concave 
X-ray spectrum, as the IPC band has a higher mean effective energy (Urry \ea
1996). These effects might explain the different energy indices obtained by us
and by Worrall \& Wilkes (1990) for FSRQ and RBL, under the same assumption of
Galactic absorption. 

As regards the Brunner \ea (1994) {\it ROSAT} results, who found $\alpha_{\rm
x} = 0.59\pm 0.19$ and $\alpha_{\rm x} = 1.36\pm 0.27$ for 8 S5 FSRQ and 5 S5
BL Lacs respectively, there the problem stems, oddly enough, from their use of
a ``complete'' (flux-limited) sample. 
It is well known that FSRQ are more luminous than BL Lacs (see e.g. Padovani
1992). Therefore, in flux-limited samples FSRQ are bound to reach higher
redshifts. For example, in the 2-Jy catalogue (Wall \& Peacock 1985; di Serego
Alighieri \ea 1994) $\langle z_{\rm BL Lacs}\rangle \simeq 0.3$ while $\langle
z_{\rm FSRQ}\rangle \simeq 1.1$. Similarly, in the 1-Jy catalogue (Stickel \ea
1994), $\langle z_{\rm BL Lacs}\rangle \simeq 0.5$ while $\langle z_{\rm
FSRQ}\rangle \simeq 1.2$. The S5 sample is no exception, so the two classes have
largely different and barely overlapping redshift distributions. In fact,
$\langle z_{\rm BL Lacs}\rangle \simeq 0.4$ while $\langle z_{\rm FSRQ}\rangle
\simeq 1.5$, with only two out of eight FSRQ having $z \le 0.77$, the
largest redshift for the BL Lacs (and these low-redshift FSRQ have $\alpha_{\rm
x} \sim 1$), while four FSRQ (i.e. 50 per cent) are at $z > 1.5$ (three at $z >
2$). Dividing the FSRQ sample of Brunner \ea in two subsamples we find that
the four FSRQ with $z > 1.5$ have $\alpha_{\rm x} = 0.05\pm0.22$, while the
four FSRQ at lower redshift have $\alpha_{\rm x} = 0.91\pm0.17$ (the latter
different from the former at the $\sim 3 \sigma$ level). Given this flattening
with redshift of the X-ray energy indices of FSRQ (particularly marked at $z >
2$: Fiore et al., in preparation), it is not surprising that Brunner \ea find
that the FSRQ in their sample have flatter spectra than their BL Lacs (which
moreover include an HBL). Our larger and
non flux-limited sample allows us to better cover the parameter space,
particularly the \ax~- $z$ plane, with a resulting large overlap between the 
redshift distributions of FSRQ and BL Lacs.  

Sambruna, Maraschi \& Urry (1996) have recently studied the multifrequency
spectra of the 1-Jy RBL, the {\it Einstein} Extended Medium Sensitivity Survey
(EMSS) X-ray selected BL Lacs (XBL), and the S5 FSRQ of Brunner \ea (1994).
Their results on the shape of the optical-to-X-ray- continua of HBL ($\simeq$
XBL) and LBL ($\simeq$ RBL) agree with those of Padovani \& Giommi (1996).
However, they find 
that FSRQ have concave only (that is \ax~$< \alpha_{\rm ox}$ for all objects)
spectra, at variance with the $\alpha_{\rm x} - \alpha_{\rm ox}$ distribution
derived in this paper, which shows that, although the majority of FSRQ have
indeed concave optical-to-X-ray- continua, a few objects with convex spectra
exist, as is the case for LBL (compare our mean value for FSRQ $\langle
\alpha_{\rm x} - \alpha_{\rm ox} = -0.28 \rangle$, which gets even higher at
lower redshifts, with their value $\langle \alpha_{\rm x} - \alpha_{\rm ox}
\rangle = -1.01$, and our Fig. 4 with their Fig. 2, taking into account that
they use the parameter $\alpha_{\rm ox} - \alpha_{\rm 
x}$). Again, this is due to the bias towards flat X-ray spectra intrinsic in
the Brunner \ea sample, discussed above. Based on these results, Sambruna \ea
(1996) have introduced a subclass of RBL, named FSRQ-like, defined by
$\alpha_{\rm x} - \alpha_{\rm ox} < -0.5$, to characterize RBL with extremely
concave X-ray spectra. As we have shown that the $\alpha_{\rm x} - \alpha_{\rm
ox}$ mean values for LBL and FSRQ are quite similar and that in fact the
majority of FSRQ has $\alpha_{\rm x} - \alpha_{\rm ox} > -0.5$, we believe
there is no need to introduce this new class of RBL. 

What are the implications of our findings for the proposed connections between
FSRQ and BL Lacs? The fact that the X-ray spectra of FSRQ and LBL are similar
might seem to be a point in favour of the microlensing hypothesis (Ostriker \&
Vietri 1985). In this picture, microlensing by stars in a foreground galaxy
could turn a distant FSRQ into a BL Lac by amplification of the relatively
compact optical continuum and consequent reduction of the equivalent widths of
the emission lines. Given that models of continuum emission in blazars suggest
relatively small sizes for the X-ray sources (e.g. Ghisellini \& Maraschi
1989), and that gravitational lensing is an achromatic process, one would then
expect the X-ray spectra of FSRQ and BL Lacs to be indistinguishable. However,
the question then rises: what about HBL? Why should there be a population of BL
Lacs with X-ray spectra steeper than both FSRQ and LBL? One would then be
forced to argue that the two classes of BL Lacs represent indeed completely
different phenomena, LBL being microlensed FSRQ and HBL being ``true'' BL
Lacs. Having two different mechanisms which produce two classes with quite
similar properties would clearly be an unappealing situation requiring some
sort of ``cosmic conspiracy''. A possible way out would be to say that, as
HBL have multifrequency spectra different from those of FSRQ, they could be
the micro-lensed version of some other class of sources, which should have
similar properties to those of FSRQ without having the same multifrequency
spectra. The existence of such a class of objects is still an open question. 
It is our view, however,  
given also the various problems of the microlensing hypothesis as an
explanation of the BL Lac phenomenon (Urry \& Padovani 1995), that the fact
that only one class of BL Lacs has X-ray spectra similar to those of FSRQ may
actually be yet one more argument against it. 

If FSRQ evolve into BL Lacs through an increase with cosmic time of the Lorentz
factor (Vagnetti \ea 1991; Vagnetti \& Spera 1994), this should affect their
X-ray spectra. The shape of the X-ray emission depends in some jet models on
the Doppler factor of the emitting material, as the bulk velocity increases
along the jet (e.g. Ghisellini \& Maraschi 1989). The net result is that in
less beamed objects synchrotron emission should dominate the X-ray band, while
the flatter inverse Compton radiation should be more important in sources with
higher Doppler factor. It then follows that, in this picture, BL Lacs should
have flatter X-ray spectra than FSRQ, in contrast with our results, which 
would suggest, at a first order, similar Doppler factors for the two classes.
It should be kept in mind, however, that this is a model-dependent conclusion
and that other parameters, beside orientation (e.g., magnetic field, jet size, 
etc.) affect the shape of the emission (see e.g. Sambruna \ea 1996). 

The main conclusions of this paper, which studies the {\it ROSAT} X-ray
spectra of more than one hundred FSRQ, can be summarized as below. 

FSRQ are characterized by energy power-law spectral indices ranging between
$\sim 0$ and $3$, with an average value $\sim 1$. No correlation is present 
between \ax~and the effective optical-X-ray spectral index, with the majority
of sources (70 per cent) displaying a concave overall spectrum (i.e. \ax~$< 
\alpha_{\rm ox}$). This is similar to what found for LBL (BL Lacs with low
frequency -- infrared/optical -- breaks in their spectra) in Paper I.
Moreover, contrary to the results obtained in previous studies, which were
however strongly affected by low spectral resolution, small number statistics
and selection biases, the X-ray spectral indices distributions of FSRQ and LBL
are indistinguishable, while both classes have X-ray spectra that are
significantly flatter than those of HBL (BL Lacs with high frequency --
UV/X-ray -- breaks in their spectra). 

These findings strongly support the hypothesis that in most FSRQ, as in LBL,
an inverse Compton component dominates the X-ray emission, and favour the idea 
that FSRQ and BL Lacs represent similar phenomena, hosted by radio galaxies of
different power. 

\section*{Acknowledgments}

We thank Fausto Vagnetti for useful discussions. This research has made use of
the BROWSE program developed by the ESA/EXOSAT Observatory and by NASA/HEASARC
and of the NASA/IPAC Extragalactic Database (NED), which is operated by the
Jet Propulsion Laboratory, California Institute of Technology, under contract
with the National Aeronautic and Space Administration.

\end